\documentclass[utf8]{frontiersFPHY} 

\setcitestyle{square} 
\usepackage{url,hyperref,lineno,microtype,subcaption}
\usepackage[onehalfspacing]{setspace}

\usepackage[T1]{fontenc}
\usepackage{color}

\def\keyFont{\fontsize{8}{11}\helveticabold }
\def\firstAuthorLast{Washiyama et al.} 
\def\Authors{Kouhei Washiyama\,$^{1,*}$, Kazuyuki Sekizawa\,$^{2,3}$}
\def\Address{$^{1}$Research Center for Superheavy elements, Kyushu University, Fukuoka, Japan \\
$^{2}$Center for Transdisciplinary Research, Institute for Research Promotion, Niigata University, Niigata, Japan \\
$^{3}$Division of Nuclear Physics, Center for Computational Sciences, University of Tsukuba, Ibaraki, Japan
}
\def\corrAuthor{Kouhei Washiyama}

\def\corrEmail{washiyama@phys.kyushu-u.ac.jp}

\begin{document}
\onecolumn
\firstpage{1}

\title[TDHF and a macroscopic aspect]{TDHF and a macroscopic aspect of low-energy nuclear reactions} 

\author[\firstAuthorLast ]{\Authors} 
\address{} 
\correspondance{} 

\extraAuth{}

\maketitle

\begin{abstract}

Time-dependent Hartree--Fock (TDHF) method
has been applied to various
low-energy nuclear reactions, such as fusion, fission,
and multinucleon transfer reactions.
In this Mini Review, we summarize recent attempts
to bridge a microscopic nuclear reaction theory, TDHF,
and 
a macroscopic aspect of nuclear reactions
through nucleus--nucleus potentials
and energy dissipation
from macroscopic degrees of freedom to microscopic ones
obtained from TDHF
in various colliding systems from light to heavy mass regions.  
  


\tiny
\keyFont{ \section{Keywords:} Heavy-ion fusion reactions, TDHF,
 nucleus--nucleus potential, energy dissipation, fusion hindrance, quasifission} 
\end{abstract}

\section{Introduction}

Time-dependent Hartree--Fock (TDHF) method has been widely used
in analyzing low-energy nuclear reactions since
Bonche and his coworkers applied TDHF to collision of slabs
in one-dimensional space 
as the first application of TDHF to nuclear physics \citep{bonche76}.
%
  Since then TDHF has been improved
%
in several respects,
e.g.,
including all terms in recent energy density functionals (EDF)
such as Skyrme \citep{umar86} and Gogny \citep{hashimoto12} functionals
and breaking symmetries
such as space (from one-dimensional to three-dimensional space).

It is well known that the coupling between
relative motions of colliding nuclei
(macroscopic degrees of freedom)
and internal excitations
of them
(microscopic degrees of freedom)
plays an important role for describing low-energy nuclear reactions
at energies around the Coulomb barrier.
To include such couplings, 
coupled-channel models \citep{reisdorf82,dasso83,balantekin98,hagino12}
have been developed and widely used.
%
TDHF automatically includes couplings between
relative
%
  motion
%
and internal excitations 
since TDHF describes the dynamics of single particles.
Moreover, TDHF provides an intuitive picture of nuclear dynamics
%
  through the time evolution
of one-body densities constructed
from single-particle wave functions in nuclei.
Recently, TDHF has been applied to
nuclear collective excitations \citep{simenel01,simenel03,nakatsukasa05,maruhn05,avez08,ebata10,stetcu11,hashimoto12,scamps13b}
and 
to nuclear reactions
such as fusion \citep{kim97,umar06a,umar09,washiyama08,washiyama09,jiang14,hashimoto16},
quasifission \citep{kedziora10,wakhle14,sekizawa16},
fission \citep{simenel14,scamps15,goddard15,bulgac16},
and
multinucleon transfer reactions \citep{simenel10,sekizawa13,scamps13,wu19,jiang20},
some of which include pairing correlations.

In this Mini Review, however,
we do not discuss the development of TDHF itself 
(see recent review articles on the development of TDHF in
Refs.~\citep{simenel12, nakatsukasa12, nakatsukasa16, simenel18, sekizawa19,stevenson19}).
Instead, we focus on a macroscopic aspect of low-energy nuclear reactions
described by TDHF.
To this end, we show various applications of
the method called ``dissipative-dynamics TDHF'' (DD-TDHF)
developed in Refs.~\citep{washiyama08,washiyama09,washiyama15}.


\section{Dissipative-dynamics TDHF}

The basic idea of DD-TDHF is to combine microscopic dynamics of nuclear reactions
described by TDHF and a macroscopic aspect of nuclear reactions
through a mapping
from microscopic TDHF evolution to a
set of
macroscopic equations
of motion.
We briefly summarize DD-TDHF by the following steps:
(1) We first solve
the
TDHF equation to obtain
time evolution of
single-particle wave functions
for nuclear reactions:
\begin{equation}
  i\hbar \frac{\partial \phi_i(t)}{\partial t}
  = \hat{h}[\rho(t)] \phi_i(t), \label{eq:tdhf}
\end{equation}
where $\phi_i(t)$ is the single-particle wave functions
with index $i$ (including spin and isospin degrees of freedom),
and $ \hat{h}[\rho(t)]$ is the single-particle Hamiltonian
as a functional of one-body density $\rho(t)$,
obtained from
an
EDF $E[\rho]$ by
an appropriate functional derivative
$\hat{h}[\rho(t)] = \delta E/\delta \rho$.
%
(2) The next step is to define macroscopic two-body dynamics from microscopic TDHF simulations.
Macroscopic two-body dynamics can be constructed
once collective coordinate is defined from TDHF simulations.
  %
  To do so in TDHF, we introduce a separation plane which
  divides the density $\rho(\boldsymbol{r},t)$ of
  a colliding system to two subsystems, $\rho_1(\boldsymbol{r},t)$ and $\rho_2(\boldsymbol{r},t)$,
  corresponding to projectile-like and target-like densities.
  This separation plane is perpendicular to the collision axis,
  and at the position where the two densities $\rho_P(\boldsymbol{r},t)$
  and $\rho_T(\boldsymbol{r},t)$ constructed from the single-particle wave functions
  initially in the projectile and in the target, respectively, cross.
(See Fig. 1 of Ref. \citep{washiyama08} for an illustrative example.)
%
We then compute the coordinate $R_i$ and
its conjugated momentum $P_i$
for each subsystem $i=1,2$ from $\rho_1(\boldsymbol{r},t)$ and $\rho_2(\boldsymbol{r},t)$.
Also, we compute the masses of the two subsystems by $m_i = P_i/\dot{R}_i$.
From these, two-body dynamics for the relative distance $R$
as a collective coordinate and its conjugated momentum $P$,
and reduced mass $\mu$ that may depend on $R$ is constructed.
(3) For the case of central collisions,
we assume that the trajectory of the two-body dynamics
obtained from TDHF
follows
a one-dimensional equation of motion for relative motions,
\begin{align}
  \frac{dR}{dt} &= \frac{P}{\mu},\label{eq:newton1} \\
  \frac{dP}{dt} &= -\frac{dV}{dR} - \frac{d}{dR} \left(\frac{P^2}{2\mu}\right)-\gamma \frac{P}{\mu},\label{eq:newton2}
\end{align}
where $V(R)$ and $\gamma(R)$ denote the nucleus--nucleus potential and friction coefficient expressing energy dissipation from the relative motion
of colliding nuclei 
to internal excitations in nuclei, respectively.
An important point is that these two quantities $V(R)$ and $\gamma(R)$
are unknown
in
 TDHF simulations.
(4) To obtain those two unknown quantities
we prepare a system of two equations from two trajectories at
slightly different energies.
Then, we solve the system of two equations at each $R$
to obtain $V(R)$ and $\gamma(R)$.
%
The details of numerical procedures for the calculations described above
can be found in Refs.~\citep{washiyama08,washiyama09,washiyama15}. 
In the following results, we used the SLy4d Skyrme EDF \citep{kim97}
without pairing interactions.


\section{Nucleus--nucleus potential and energy dissipation}

\subsection{Light and medium-mass systems}\label{sec:light}

In light and medium-mass systems, whose charge product $Z_1Z_2$ is
smaller than $\approx 1600$,
it is known that fusion occurs once two nuclei contact
each other after passing the Coulomb barrier.
Indeed, TDHF simulations
for head-on collisions at energies above the Coulomb barrier
lead to fusion, keeping a compound system compact 
for sufficiently long time. 
We first provide selected results of nucleus--nucleus potential
and energy dissipation obtained from DD-TDHF
and discuss their properties.

\begin{figure}[t]
\begin{center}
  \includegraphics[width=0.98\linewidth]{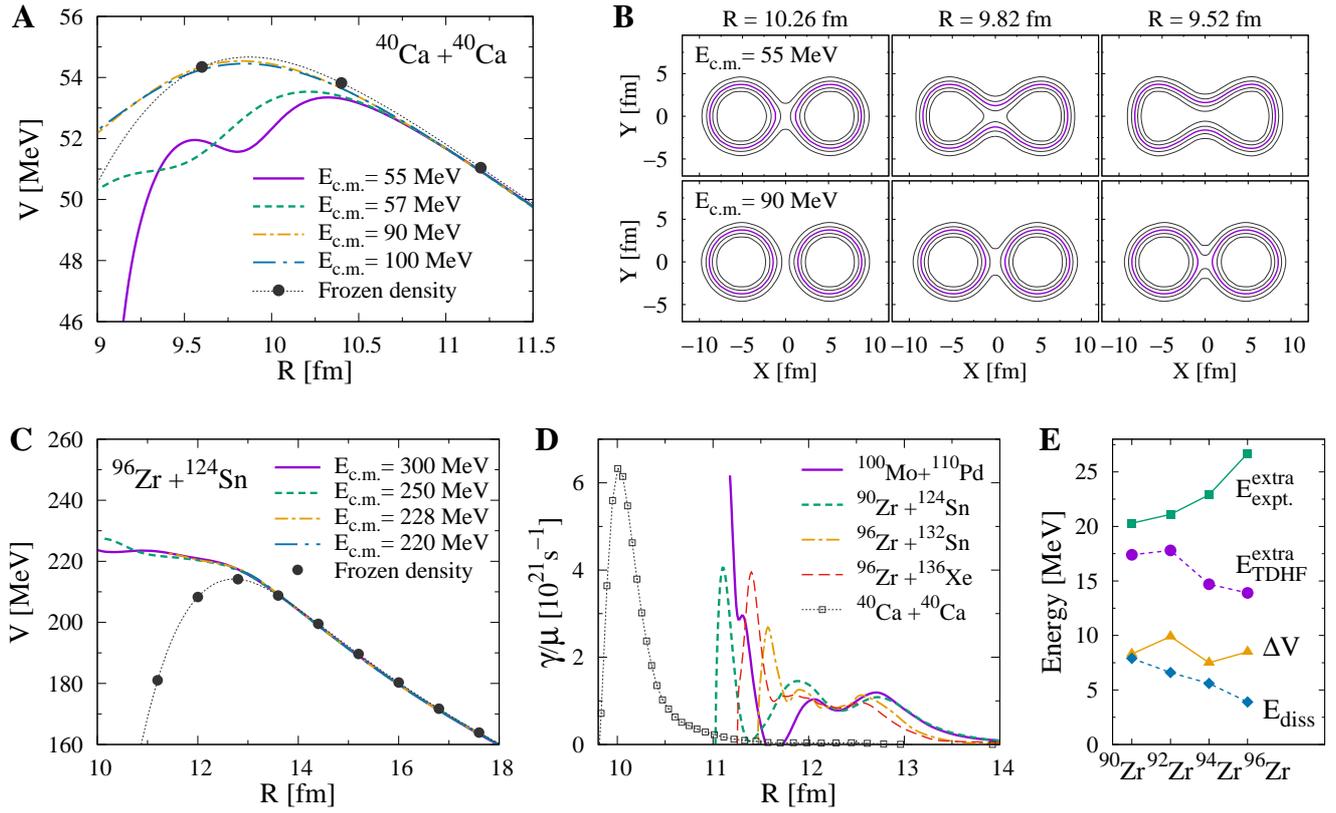}
\end{center}
\caption{
  \textbf{(A)} Nucleus--nucleus potentials denoted by lines obtained by DD-TDHF at different energies
  and by the frozen-density approximation by filled circles with the dotted line
  as a function of $R$ in $^{40}\text{Ca}$ + $^{40}\text{Ca}$.
  \textbf{(B)} Contour plots of density profile $\rho(X,Y,0)$ obtained from
  TDHF at $E_{\text{c.m.}}=55$\,MeV (upper panels) and at $E_{\text{c.m.}}=90$\,MeV (lower panels), at $R=10.26$\,fm (left panels), $9.82$\,fm (middle), and $9.52$\,fm (right) in $^{40}\text{Ca}$ + $^{40}\text{Ca}$.
  The isodensities (contour lines) are plotted at each $0.025$\,$\text{fm}^{-3}$.
  \textbf{(C)} Same as \textbf{(A)} but for $^{96}\text{Zr}$ + $^{124}\text{Sn}$.
  \textbf{(D)} Friction coefficient divided by reduced mass, $\gamma/\mu$, from DD-TDHF.
  \textbf{(E)} Extra-push energy from experiments ($E^{\text{extra}}_{\text{expt.}}$)
  and from TDHF ($E^{\text{extra}}_{\text{TDHF}}$) together with
  potential increase $\Delta V$ and dissipated energy $E_{\text{diss}}$
  for $^{90,92,94,96}\text{Zr}$ + $^{124}\text{Sn}$ (see text for detail).
  \textbf{(A)} and \textbf{(B)} adopted from \citep{washiyama08},  
  \textbf{(C)} and \textbf{(D)} adopted from \citep{washiyama15},
  and  \textbf{(E)} adopted from \citep{washiyama15} with slight change
  with permission from APS and SciPris.
}\label{fig:pot_density}
\end{figure}

In Fig. \ref{fig:pot_density}A,
we show obtained nucleus--nucleus potentials
as a function of relative distance $R$
near the Coulomb barrier radius for $^{40}\text{Ca}$ + $^{40}\text{Ca}$.
The lines show the nucleus--nucleus potentials
at different center-of-mass energies ($E_{\text{c.m.}}$) by DD-TDHF,
while the filled circles show 
the potential obtained by the frozen-density approximation,
  where the energy of the system is calculated with the same EDF
  except that the dynamical effect during the collision is neglected
  and the density of each fragment is fixed to be its ground-state one.
  Moreover, in the frozen-density approximation, the Pauli principle is
  neglected between nucleons in the projectile and in the target,
  leading to worse approximation as the overlap of projectile and target nuclei becomes significant.
%
Important remarks from this figure are:
(1) Potentials obtained at higher energies ($E_{\text{c.m.}}=90$, $100$\,MeV)
agree with the frozen-density one,
indicating the convergence of the potentials obtained by DD-TDHF
at higher energies.
(2) DD-TDHF potentials express an $E_{\text{c.m.}}$ dependence
  at lower energies $E_{\text{c.m.}}=55$, $57$\,MeV.
%
(3) The height of DD-TDHF potential decreases
with decreasing $E_{\text{c.m.}}$.
  The Coulomb barrier height decreases from $\approx 54.5$\,MeV at $E_{\text{c.m.}}=90$, $100$\,MeV of DD-TDHF and of the frozen-density approximation
  to $\approx 53.4$\,MeV at $E_{\text{c.m.}}=55$\,MeV of DD-TDHF.
The above remarks can be understood by the dynamical
reorganization of the TDHF density profile of each TDHF trajectory.
Figure \ref{fig:pot_density}B shows
the TDHF density $\rho(x,y,z=0,t)$ at each $R$
for $E_{\text{c.m.}}=55$ (top panels) and $90$\,MeV (bottom panels).
At $E_{\text{c.m.}}=90$\,MeV, the shape of each ${}^{40}$Ca density
keeps its shape spherical,
while at $E_{\text{c.m.}}=55$\,MeV the shape of  each ${}^{40}$Ca density
deviates from its ground-state spherical shape as $R$ becomes smaller.
This is a dynamical reorganization of density during fusion reactions.
This dynamical reorganization changes the shape of each nucleus
when two nuclei approach sufficiently, then reduces the height of the
nucleus--nucleus potential obtained by DD-TDHF.
This dynamical reduction of the nucleus--nucleus potential
is seen in various light- and medium-mass systems
in Ref.~\citep{washiyama08}.

We would like to note that, 
in the density-constrained TDHF (DC-TDHF) method \citep{umar06a},
%
in which constrained Hartree--Fock calculation is performed
to obtain the nucleus--nucleus potential under the condition that
the density is constrained to the density obtained from TDHF at each time,
similar $E_{\text{c.m.}}$ dependence of nucleus--nucleus potentials
is seen in various colliding systems
reported, e.g., in Refs. \citep{umar09,umar10,oberacker10,umar14}.
Moreover, in the $^{40}\text{Ca}$ + $^{40}\text{Ca}$ system,
we find no significant difference in the potential extracted by DD-TDHF and 
the one by DC-TDHF \citep{umar14}.

\subsection{Heavy systems}

Contrary to light and medium-mass systems described
in Sec.~\ref{sec:light},
it was experimentally observed that fusion probability at energies
near the Coulomb barrier is strongly hindered in heavy systems
($Z_1 Z_2 \ge 1600$) \citep{gaggeler84,sahm84}.
%
The main origin of this hindrance
has been considered as the presence of
the quasifission process, where a composite system
of
two colliding nuclei
reseparates before forming an equilibrated compound nucleus.
%
%
%
This fusion hindrance indeed has been
observed
in TDHF
e.g., in Refs. \citep{simenel12,washiyama15,simenel09,guo12}.
Namely, TDHF simulations
for head-on collisions at energies above the Coulomb barrier
lead to touching configuration of a composite system,
and then to reseparation after
a while (several to tens of zeptoseconds).
%
In Ref.~\citep{washiyama15},
  the extra-push energy $E^{\text{TDHF}}_{\text{extra}} = E^{\text{TDHF}}_{\text{thres}} - V^{\text{FD}}_{B}$
in TDHF was systematically obtained in heavy systems,
where $E^{\text{TDHF}}_{\text{thres}}$ and $V^{\text{FD}}_{B}$ denote the fusion threshold energy
above which fusion occurs in TDHF 
and the Coulomb barrier energy obtained in the frozen-density approximation, respectively.
We show in Fig. \ref{fig:pot_density}E
extra-push energies in TDHF for $^{90,92,94,96}\text{Zr}$ + $^{124}\text{Sn}$,
compared with those deduced from experimental data, $E^{\text{extra}}_{\text{expt.}}$,
taken from \citep{schmidt91}, where the Bass barrier $V_{\text{Bass}}$ \citep{bass74} was employed as
the Coulomb barrier height.
We found that the difference between $V^{\text{FD}}_{B}$ and $V_{\text{Bass}}$
in $^{90,92,94,96}\text{Zr}$ + $^{124}\text{Sn}$ is at most $\approx 1$\,MeV.
These obtained extra-push energies in TDHF reasonably reproduce observations.


One may think why the fusion hindrance in heavy systems appears in both
experiments and TDHF simulations.
In Ref.~\citep{washiyama15}, we address this question
and analyze where finite extra-push energies arise.
For the analysis,
we first derive the nucleus--nucleus potential and energy dissipation
by DD-TDHF
because we think that these two quantities are strongly related to
the appearance of finite extra-push energy.
In Fig.~\ref{fig:pot_density}C,
we show an example of nucleus--nucleus potentials extracted in heavy systems,
which is for the $^{96}\text{Zr}$ + $^{124}\text{Sn}$ system
for three different energies in DD-TDHF and the frozen-density one.
One can clearly see the difference between the potentials
in $^{40}\text{Ca}$ + $^{40}\text{Ca}$ (Fig. \ref{fig:pot_density}A)
and $^{96}\text{Zr}$ + $^{124}\text{Sn}$ (Fig. \ref{fig:pot_density}C):
the potentials
in $^{96}\text{Zr}$ + $^{124}\text{Sn}$
extracted by DD-TDHF monotonically
increases as the relative distance decreases
while the potentials in $^{40}\text{Ca}$ + $^{40}\text{Ca}$ and
by the frozen-density approximation in $^{96}\text{Zr}$ + $^{124}\text{Sn}$
show a barrier structure at a certain relative distance.
We have observed monotonic increase in potential
in other heavy systems \citep{washiyama15}.
We consider the increase in potential in heavy systems
as the transition from two-body dynamics of colliding nuclei
to one-body dynamics of a composite system with strong overlap of
the densities of
colliding nuclei in TDHF
and as the appearance of the conditional saddle point
inside the Coulomb barrier in heavy systems 
\citep{swiatecki81,swiatecki82,bjornholm82,swiatecki05}.

We would like to
%
%
note here that this is different property 
from the one obtained from the DC-TDHF method in 
the same colliding system in Ref.~\citep{oberacker10}.
This difference comes from a different interpretation
of the nucleus--nucleus potential between the two methods.
%
In the DC-TDHF method energy minimization is
carried out at a given density of a system obtained from TDHF
to deduce a nucleus--nucleus potential that eliminates
internal excitations in this system.
In the DD-TDHF method the potential
is deduced under the assumption that TDHF evolution
is reduced to a one-dimensional equation of motion for relative motion.
We consider that 
the DD-TDHF potential can include a part of
the DC-TDHF internal excitation energy.
We make a comment on the origin of the difference
between the two potentials in the following:
In heavy systems with larger Coulomb replusion,
larger overlap of projectile and target densities
during a collsion in TDHF is achieved at a short relative distance.
In TDHF, diabatic level crossings can occur more in larger overlap region,
leading to a part of internal excitations
and to a transition from two-body to one-body picture of a system.
This part of internal excitations is interpreted as potential energy in DD-TDHF,
while this is treated as excitation energy in DC-TDHF.
%
In the DC-TDHF method, 
the flattening of the potential at short distances inside
the Coulomb barrier radius is seen in heavier systems
leading to the synthesis of superheavy elements in Ref.~\citep{umar10}.

In Fig. \ref{fig:pot_density}D,
reduced friction coefficient ($\gamma/\mu$),
the friction coefficient
divided by the reduced mass extracted from Eq. \eqref{eq:newton1},
are plotted for selected systems.
The friction coefficient increases as $R$ decreases,
and shows oscillations in heavy systems. 
We consider that
the fact that the friction coefficient becomes negative
indicates
breakdown
of the assumption that
the TDHF trajectory follows a macroscopic one-dimensional equation of motion
for relative motion of a two-body colliding system.

Finally we consider the origin of the fusion hindrance in heavy systems
through the analysis with DD-TDHF.
As mentioned above,
nucleus--nucleus potential and energy dissipation are
main contribution to the appearance of finite extra-push energy.
We evaluate
the potential increase
at short distances
and the accumulated dissipation energy from the friction coefficient using
the formula \citep{washiyama15},
\begin{equation}
  E_{\text{diss}}(t) = \int_{0}^t dt' \gamma[R(t')] \dot{R}(t')^2 ,
\end{equation}
up to time $t$ when
the kinetic energy of the relative motion
of the system is completely dissipated.
In Fig.~\ref{fig:pot_density}E,
we also show the contribution of
potential increase $\Delta V$ and
dissipated energy $E_{\text{diss}}$ to the extra-push energy
in the $^{90,92,94,96}\text{Zr}$ + $^{124}\text{Sn}$ systems.
The result $\Delta V > E_{\text{diss}}$
indicates that the potential increase
is a main origin for the appearance of the finite extra-push energy,
i.e., fusion hindrance.
Though the energy dissipation is known to play an important role
in this fusion hindrance, it is not sufficient to explain the
amount of the extra-push energy in the analysis with the DD-TDHF method.

\subsection{Off-central collisions}


So far, the applications of the DD-TDHF method has been limited to
central collisions. Here we discuss a possible extension of the method
to off-central collisions. Regarding $(R,P)$ and $(\varphi,L)$ as sets
of canonical coordinates, where $\varphi$ represents a rotation angle
of the colliding system in the reaction plane and $L=\mu R^2\dot\varphi$
is the angular momentum of the relative motion, we obtain a set of
macroscopic
equations of motion:
\begin{eqnarray}
&&\frac{{\rm d}R}{{\rm d}t} = \frac{P}{\mu},\label{Eq:dRdt}\\[1mm]
&&\frac{{\rm d}\varphi}{{\rm d}t} = \frac{L}{\mu R^2},\label{Eq:dthetadt}\\[1mm]
&&\frac{{\rm d}P}{{\rm d}t} = -\frac{{\rm d}V}{{\rm d}R}
+ \frac{1}{2}\biggl( \frac{P^2}{\mu^2} + \frac{L^2}{\mu^2R^2} \biggr) \frac{{\rm d}\mu}{{\rm d}R}
+ \frac{L^2}{\mu R^3}
- \gamma_R\frac{P}{\mu},\label{Eq:dPdt}\\[1mm]
&&\frac{{\rm d}L}{{\rm d}t} = -\gamma_\varphi\frac{L}{\mu}.\label{Eq:dLdt}
\end{eqnarray}
Here, $\gamma_R(R)$ and $\gamma_\varphi(R)$ denote the radial and tangential
(or ``sliding") friction coefficients, respectively, where
the former already appeared in Eq.~\eqref{eq:newton2}, the case of central collisions, and
the latter governs the
angular momentum dissipation [cf. Eq.~(\ref{Eq:dLdt})].

At first sight, there are three unknown quantities:
the nucleus--nucleus potential $V$,
the radial friction coefficient $\gamma_R$, and the tangential friction coefficient $\gamma_\varphi$.
However, since time evolution of $\varphi(t)$ and $L(t)$ can be obtained from TDHF,
a single TDHF simulation
already
provides the tangential friction coefficient by
\begin{equation}
\gamma_\varphi(R)=-\mu(t)\frac{\dot{L}(t)}{L(t)}.
\label{Eq:gamma_theta}
\end{equation}
Thus, there are only two unknown quantities in Eqs.~(\ref{Eq:dRdt})--(\ref{Eq:dLdt}),
i.e. $V(R)$ and $\gamma_R(R)$, and we can apply the same procedure applied for
central collisions.

\begin{figure}[t]
\begin{center}
\includegraphics[width=16cm]{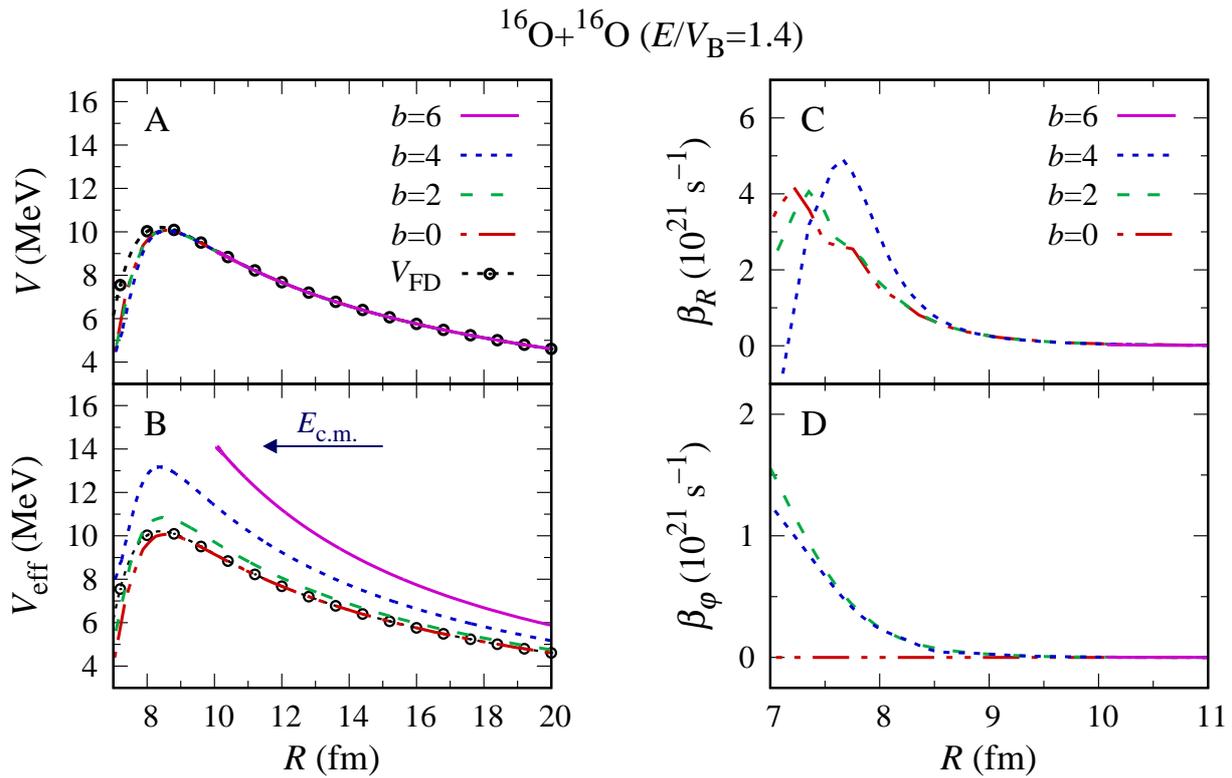}
\caption{
Results of DD-TDHF calculations for the $^{16}$O+$^{16}$O reaction
at $E/V_B=1.4$ at various impact parameters. The nucleus--nucleus
potentials and the effective potential $V_{\rm eff}(R)=V(R)+L^2/2\mu R^2$
are shown in \textbf{(A)} and \textbf{(B)}, respectively, as a function of
the relative distance, $R$. The reduced radial friction coefficients, $\beta_R=
\gamma_R/\mu$, are shown in \textbf{(C)}, while the reduced tangential
friction coefficients, $\beta_\varphi=\gamma_\varphi/\mu$, are shown in
\textbf{(D)}.
}
\label{fig:off-central}
\end{center}
\end{figure}

In Fig.~\ref{fig:off-central}, we show the results for the $^{16}$O+$^{16}$O reaction
at $E/V_{\rm B}=1.4$ including off-central collisions, as an illustrative example. In Fig.~\ref{fig:off-central}A,
the nucleus--nucleus potential is shown as a function of the relative distance, $R$. We also show
the potential in the frozen-density approximation by open circles, for comparison. Figure~\ref{fig:off-central}A
clearly shows that the method provides almost identical nucleus--nucleus potentials $V(R)$ irrespective
of the impact parameters. In Fig.~\ref{fig:off-central}B, the effective potential $V_{\rm eff}(R)$,
the sum of nuclear, Coulomb, and centrifugal potentials, is shown. It can be seen that, for $b=6$~fm,
the closest distance is achieved at around $R=10$~fm, at which the effective potential coincides
with the incident relative energy. In Figs.~\ref{fig:off-central}C and \ref{fig:off-central}D, the
reduced radial and tangential friction coefficients, $\beta_R=\gamma_R/\mu$ and $\beta_\varphi
=\gamma_\varphi/\mu$, are shown as a function of the relative distance. We found no significant
dependence of the friction coefficients on the impact parameters in this system. In this way,
this approach enables us to access the angular momentum dissipation mechanism and
a systematic calculation is in progress.

  Note that non-central effects on nucleus--nucleus potentials
  and effective mass parameters in fusion reactions
  have been studied in TDHF and DC-TDHF in Ref. \citep{jiang14}.
  It is interesting to make detailed comparison between those and
  our DD-TDHF in a future work.

\section{Summary}

The
macroscopic aspect of TDHF dynamics for low-energy
nuclear reactions at energies near the Coulomb barrier 
was
discussed within the DD-TDHF method.
We showed that the dynamical reorganization of single-particle
wave functions inside the colliding nuclei
affects the macroscopic nucleus--nucleus potential
that leads to dynamical reduction of the potential
around the Coulomb barrier radius in light- and medium-mass systems.
In heavy systems, 
the dynamical reorganization leads to the fusion hindrance,
increase in potential compared with
the potential obtained from the frozen-density approximation
in which the dynamical reorganization effect is neglected.
By extending the DD-TDHF method to off-central collisions,
the tangential friction coefficient
was extracted in the $^{16}$O+$^{16}$O reaction
in addition to the nucleus--nucleus potential and the radial friction.
As expected, the nucleus--nucleus potentials do not show
a significant dependence of the initial angular momentum.
The strength of the tangential friction is in the same order
of magnitude as the radial one.
From this extension, one can access the mechanism of
angular momentum dissipation from microscopic reaction models.
Possible future extension would be
a systematic study of angular momentum dissipation mechanism
in various systems, especially in heavy systems
to address the fusion hindrance problem.
Another possible extension would be 
a systematic study of collisions with deformed nuclei.
It is interesting to study an orientation effect, a dependence of an angle
between the collision axis and the principle axis of a deformed nucleus,
on the nucleus--nucleus potential and the friction coefficient.
It would be important to investigate
how orbital angular momentum dissipation couples to a rotation of
deformed nucleus during collision.

\section*{Conflict of Interest Statement}

The authors declare that the research was conducted in the absence of any commercial or financial relationships that could be construed as a potential conflict of interest.

\section*{Author Contributions}

The manuscript was prepared by the authors.


\section*{Funding}

The work was supported in part by QR Program of Kyushu University,
by JSPS-NSFC Bilateral Program for Joint Research Project
on Nuclear mass and life for unravelling mysteries of r-process,
and
%
by JSPS Grant-in-Aid for Early-Career Scientists No. 19K14704.


\section*{Acknowledgments}
The authors acknowledge
Denis Lacroix and Sakir Ayik for the collaboration to this work.
%
This work used computational resources of the Oakforest-PACS Supercomputer System
provided by Multidisciplinary Cooperative Research Program in Center for Computational
Sciences (CCS), University of Tsukuba (Project ID: NUCLHIC), and computational
resources of the HPCI system (Oakforest-PACS) provided by Joint Center for
Advanced High Performance Computing (JCAHPC) through the HPCI System
Project (Project ID: hp190002).



\bibliographystyle{frontiersinHLTHFPHY} 

\end{document}